\documentclass[aps,10pt,reprint,groupedaddress,superscriptaddress, prx]{revtex4-2}

\usepackage{amssymb}
\usepackage{graphicx}
\usepackage{amsmath}
\usepackage{multirow}
\usepackage{blkarray}
\usepackage{nicematrix}
\usepackage{tikz}
\NiceMatrixOptions{
code-for-first-row = \color{black} ,
code-for-last-row = \color{black} ,
code-for-first-col = \color{black} ,
code-for-last-col = \color{black}
}
\usepackage[bookmarks=true,colorlinks,citecolor=blue,urlcolor=blue]{hyperref}
\usepackage{braket}
\usepackage{babel}
\usepackage{blindtext}
\usepackage{soul}
\usepackage[compat=1.1.0]{tikz-feynman}
\usepackage[normalem]{ulem}
\usepackage{physics}

\usepackage{mathtools}

\definecolor{mypurp}{rgb}{0.35, 0, 0.7}

\usepackage{amsthm}
\usepackage{txfonts}

\theoremstyle{definition}

\begin{document}
\def\papertitle{{Simulating 2D topological quantum phase transitions on a digital quantum computer}}
\newcommand{\TUM}{\affiliation{Technical University of Munich, TUM School of Natural Sciences, Physics Department, 85748 Garching, Germany}}
\newcommand{\MCQST}{\affiliation{Munich Center for Quantum Science and Technology (MCQST), Schellingstr. 4, 80799 M{\"u}nchen, Germany}}
\newcommand{\USH}{\affiliation{Department of Physics and Astronomy, University of California at Riverside, Riverside, California 92521, USA}}

\title{\papertitle}

\author{Yu-Jie Liu}\TUM\MCQST
\author{Kirill Shtengel}\USH
\author{Frank Pollmann}\TUM\MCQST

\begin{abstract}
   Efficient preparation of many-body ground states is key to harnessing the power of quantum computers in studying quantum many-body systems. In this work, we propose a simple method to design exact linear-depth parameterized quantum circuits which prepare a family of ground states across topological quantum phase transitions in 2D. We achieve this by constructing ground states represented by isometric tensor networks (isoTNS), which form a subclass of tensor network states that are efficiently preparable. 
    By continuously tuning a parameter in the wavefunction, the many-body ground state undergoes quantum phase transitions, exhibiting distinct 2D quantum phases. We illustrate this by constructing an isoTNS path with bond dimension $D = 2$ interpolating between distinct symmetry-enriched topological (SET) phases. At the transition point, the wavefunction is related to a gapless point in the classical six-vertex model. Furthermore, the critical wavefunction supports a power-law correlation along one spatial direction while remaining long-range ordered in the other spatial direction. We provide an explicit parametrized local quantum circuit for the path and show that the 2D isoTNS can also be efficiently simulated by a holographic quantum algorithm requiring only an 1D array of qubits. 
\end{abstract}

\maketitle

\begin{figure}[t]
    \centering
    \includegraphics{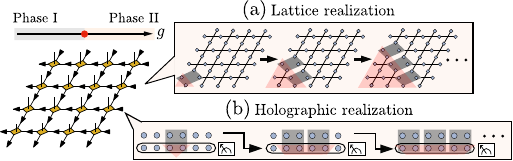}
    \caption{We construct parametrized isoTNS for 2D quantum phase transitions crossing some quantum critical point. We propose two different schemes to realize the states on a gate based quantum computer: (a) The parametrized isoTNS can be directly realized by a linear-depth sequential quantum circuit on a 2D lattice of qubits, which creates the full entangled wavefunction (see Sec.~\ref{sec:circ}). The entangled qubits are shaded by red. (b) by exploiting the causal structure of the sequential circuit, the 2D isoTNS can also be efficiently simulated using a holographic quantum algorithm which requires only an 1D array of qubits (discussed in Sec.~\ref{sec:holo}). The qubit array on the top encodes the boundary of the partially prepared isoTNS wavefunction.}
    \label{fig:overview}
\end{figure}

\section{Introduction}
The search for exotic quantum phases of matter is a central theme in condensed matter physics. While the last decades have witnessed tremendous progress in our theoretical understanding of topologically ordered phases, physical realization of topological phases remains a significant challenge, with the fractional quantum Hall effect~\cite{tsui:1982} representing one of the few unambiguous examples in the solid state. The advent of programmable quantum hardware has opened up unprecedented avenues for accessing novel quantum states. Recently, breakthroughs were made in the realization of topologically ordered states using Rydberg simulators~\cite{semeghini:2021,verresen:2021}, superconducting qubits~\cite{Satzinger2021,Liu:2022} and trapped ions~\cite{iqbal:2023,tant:2023}. While these realizations focused on specific topological states, the important question of realizing the topological quantum phase transitions, i.e., transitions that cannot be detected by any local order parameters, is more challenging. Progress has been made in one-dimensional (1D) symmetry-protected topological (SPT) systems by exploiting the correspondence between sequential quantum circuits and matrix-product states (MPS)~\cite{schon:2005}. The exact ground states across SPT phase transitions can be represented by a parameterized MPS with a finite bond dimension and can be physically realized using its efficient quantum circuit representation~\cite{wolf:2006, jones:2021, smith:2022}. General two-dimensional (2D) tensor-network states (TNS) with a finite bond dimension can describe exact ground states across various quantum phase transitions---including some critical states with a power-law correlation~\cite{Verstraete:2006,xu:2018,zhu:2019,xu:2020,zhang:2020a,Xu:2021,Xu:2022, Haller:2023}. However, general 2D TNS cannot be efficiently prepared on a quantum computer~\cite{schuch:2007}.  

A direct approach to realize a 2D quantum phase transition on a quantum computer would be to parametrize the quantum circuits in order to efficiently prepare the fixed-point ground states and then tune the parameters away from the fixed points. However, with this approach it is difficult to preserve global symmetries of the system when deforming the circuit. A more subtle issue appears if the state has long-range entanglement: a small deformation of the local gates can result in a non-local perturbation to the wavefunction, potentially changing its long-range entanglement properties~\cite{shukla:2018}. This is particularly problematic when the state has an intrinsic topological order characterized by long-range entanglement. While these problems pose a significant challenge in working directly with quantum-circuit representation, they can be conveniently addressed in the tensor-network representation: A global on-site symmetry can be preserved by enforcing local conditions on the tensor~\cite{cirac:2021}. The stability of the long-range entanglement can be ensured by enforcing the correct virtual symmetries on the local tensors~\cite{williamson:2016,Williamson:2017,shukla:2018}.

Motivated by these considerations, we use guidance from tensor networks and investigate topological quantum phase transitions between states \emph{exactly} representable by 2D isometric tensor-network states (isoTNS)~\cite{zaletel:2020}. IsoTNS form a subclass of 2D TNS with an additional isometry condition and can represent a large class of gapped quantum phases~\cite{soejima:2020}. 
The isometry condition establishes a 2D analogue of the canonical form in 1D MPS and directly leads to the correspondence between isoTNS and linear sequentially generated quantum circuits~\cite{zaletel:2020, wei:2022} (see also Appendix~\ref{sm:isotns_circ} for a brief discussion). The isoTNS have the advantages of both the tensor-network and efficient quantum-circuit presentation; they can therefore serve as an ideal starting point for exploring the efficient realization of ground states across 2D quantum phase transitions, going beyond realizing specific fixed points.

 Our goal here is to demonstrate the use of isoTNS as a general tool for designing efficient quantum circuits across quantum phase transitions (see Fig.~\ref{fig:overview}). We propose a simple ``plumbing'' method to construct isoTNS such that the coefficients in the wavefunction can be associated with the Boltzmann weights of certain 2D classical partition functions. By introducing an internal parameter, the system can be deformed continuously from one phase to another phase via a quantum phase transition. We illustrate this method by constructing a quantum phase transition in the ground states between symmetry-enriched topological (SET) phases, where the system has an intrinsic $\mathbb{Z}_2$ topological order enriched by an anti-unitary $\mathbb{Z}_2^T$ symmetry. We discuss the properties of the ground state both away from and at the transition point. We then explicitly show that apart from the direct lattice realization of the isoTNS using quantum circuits, the 2D isoTNS can also be efficiently simulated using a holographic quantum algorithm (see Fig.~\ref{fig:overview}). This leads to a minimal experimental proposal for simulating and detecting an SET phase transition on current quantum hardware.

\begin{figure}[t]
    \centering
    \includegraphics{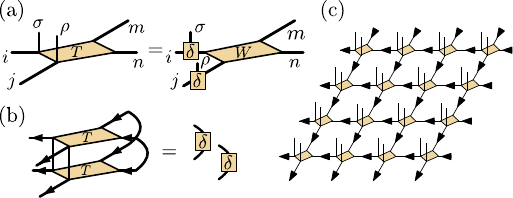}
    \caption{Construction of isoTNS: Each yellow node labels a tensor. (a) Each tensor has two physical legs ($\sigma$ and $\rho$) and four virtual legs $i,j,m,n$. The local tensor $T$ is a product of some matrix $W$ and two $\delta$-tensors. The $\delta$-tensor is 1 when all the legs are equal, and zero otherwise. (b) The isometry condition. The contraction of the tensor $T$ with its complex conjugate (depicted with flipped physical legs) yields two two-leg $\delta$-tensors. The arrows indicate the direction of the contraction.
   (c) The wavefunction is obtained by contracting the virtual legs of the tensors. The physical legs of the tensors correspond to the two physical degrees of freedom on the links. 
   }
    \label{fig:isotns}
\end{figure}

\section{Isometric Tensor Networks and Classical Partition Functions}
We focus our discussion on 2D systems, the generalization of the construction to arbitrary higher dimensions is straightforward.
We begin by briefly reviewing the concept of 2D TNS (Fig.~\ref{fig:isotns}). Consider a 2D spin system on a square lattice with local Hilbert space dimension $d$ and each spin is located on the edges of the square lattice. A 2D TNS can be defined via a rank-6 local tensor $T^{\sigma\rho}_{ijmn}$ at each vertex, where $\sigma$ ($\rho$) represents the spin degree of freedom on the left (bottom) edge connected to the vertex. The legs $i,j,m,n$ label the virtual degrees of freedom (Fig.~\ref{fig:isotns}a). The dimension of the virtual legs is referred to as the bond dimension. For a translationally invariant system with $N$ spins, the wavefunction is obtained by contracting the neighbouring virtual legs of all the tensors
\begin{equation}
    \ket{\psi} = \sum_{\sigma_1,\cdots,\sigma_N}\text{tTr}\left(\{T^{\sigma_1\sigma_2}, \cdots, T^{\sigma_{N-1}\sigma_N}\}\right)\ket{\sigma_1,\cdots,\sigma_N},
\end{equation}
where tTr denotes the tensor contraction. 

Motivated by the plasma analogy~\cite{laughlin:1983}, we can encode classical partition functions in the TNS with a finite bond dimension~\cite{Verstraete:2006}  so that the squared norms of the coefficients in the wavefunction are identified with the Boltzmann weights in the classical partition function. The thermal phase transition in the classical system is thus mapped onto a quantum phase transition, whereby classical criticality corresponds to the critical correlations in the wavefunction. In what follows, we restrict this class of tensor networks to isoTNS.
To achieve this, we impose the following conditions on the local tensors:

(i)
The local tensor $T^{\sigma\rho}_{ijmn}$ can be decomposed as
\begin{equation}\label{eq:plumb}
    T^{\sigma\rho}_{ijmn} = \sum_{i',j'}\delta^{\sigma}_{ii'}\delta^{\rho}_{jj'}W_{i'j'mn},
\end{equation}
where $W_{i'j'mn}$ is a $d^2\times d^2$ matrix and  $\delta^{\sigma}_{ab}$ denotes a ``plumbing'' $\delta$-tensor such that $\delta^{\sigma}_{ab} = 1$ if $\sigma = a = b$, and zero otherwise. This relation is depicted in Fig.~\ref{fig:isotns}a. This condition relates the quantum wavefunction to a classical partition function in the following way: Tensor $\delta^{\sigma}_{ab}$ makes the virtual legs equivalent to the physical degrees of freedom in the quantum system. As a result, the probabilities for the spin configurations at each vertex in the tensor-network wavefunction are encoded in a local tensor $R_{ijmn} = |W_{ijmn}|^2$, where $R$ is a matrix of weights for different configurations of spins $i,j,m,n$ around the vertex. The transfer matrix of the isoTNS is thus the same as the transfer matrix for a classical partition function contracted from the local weight matrix (see Appendix~\ref{sm:qc_map}). 

(ii)
We enforce the isometry condition on the local tensor $T^{\sigma\rho}_{ijmn}$. More precisely, we require
\begin{equation}\label{eq:isometry}
    \sum_{\sigma,\rho,m,n} \left(T^{\sigma\rho}_{ijmn}\right)^*T^{\sigma\rho}_{i'j'mn} = \delta_{ii'}\delta_{jj'}.
\end{equation}
This is pictorially shown in Fig.~\ref{fig:isotns}b. The subset of TNS satisfying this condition is isoTNS. For the plumbed isoTNS satisfying Eq~\eqref{eq:plumb}, the 2D isometry condition Eq.~\eqref{eq:isometry} is satisfied if and only if
\begin{equation}\label{eq:norm}
    \sum_{m,n} |W_{ijmn}|^2 = 1,\ \forall i,j.
\end{equation}

For a given bond dimension, the $W$-matrix representing the plumbed isoTNS forms a finite dimensional manifold. Our strategy is to search for continuous paths within this manifold that connect between ground states having different quantum phases. It is worth mentioning that in 1D the plumbed isoTNS is a subclass of the canonical form for 1D matrix-product states (MPS). Simple examples of quantum phase transitions previously known in MPS~\cite{wolf:2006} can also be constructed using the plumbing construction (see Appendix~\ref{sm:mps_plumb} for details). 

\section{A Continuous IsoTNS Path Between Symmetry-Enriched Topological Phases Crossing a Quantum Critical Point}\label{sec:qpt}
In this section, we illustrate the plumbing method by constructing a minimal example of an isoTNS path across a topological quantum phase transition.
Let us consider a spin-1/2 system where each spin is encoded by a qubit with the Pauli basis such that $Z\ket{0} = \ket{0}$ and $Z\ket{1} = -\ket{1}$. The physical leg of the tensor has dimension $d = 2$; Eq.~\eqref{eq:plumb} implies that the plumbed isoTNS has bond dimension $D = d=2$. The toric code ground state with $\mathbb{Z}_2$ topological order~\cite{kitaev:2003} naturally falls into this family of isoTNS, with 
\begin{equation}
W^{(\text{TC})} = \begin{pNiceMatrix}[first-row,last-col]
 \ket{00}   & \ket{01} & \ket{10} & \ket{11} &        \\
\frac{1}{\sqrt{2}}   & 0   & 0   & \frac{1}{\sqrt{2}}   & \ \ket{00} \\
0   & \frac{1}{\sqrt{2}}   & \frac{1}{\sqrt{2}}   & 0   & \ \ket{01} \\
0   & \frac{1}{\sqrt{2}}   & \frac{1}{\sqrt{2}}   & 0   & \ \ket{10} \\
 \frac{1}{\sqrt{2}}   & 0 & 0 & \frac{1}{\sqrt{2}} & \ \ket{11}       \\
\end{pNiceMatrix},
\end{equation}
where the indices label the legs $i.j,m,n$ in $W_{ijmn}$.
If we view $\ket{1}$ as occupied by a string and  $\ket{0}$ as empty, the eight non-zero entries in $W$ are exactly the eight vertex configurations with no broken strings. The resulting wavefunction is an equal-weight superposition of all the closed-loop configurations as we expect from the toric code ground state. 

To obtain an isoTNS away from the fixed point, we deform the entries in the $W$-matrix while preserving the isometry condition in Eq.~\eqref{eq:norm}. In order to obtain a quantum phase transition, the deformed isoTNS should preserve the same symmetry as the fixed point (see Appendix~\ref{sm:sym} for more discussion). 
Consider the following continuous path of matrix $W(g)$ for $g\in[-1, 1]$,
\begin{equation}\label{eq:set_path}
    W(g)= \left(\begin{matrix}
        \frac{1}{\sqrt{1+|g|}} & 0 & 0 &  \text{sign}(g)\sqrt{\frac{|g|}{1+|g|}}\\
        0 &  \frac{1}{\sqrt{2}} &  \frac{1}{\sqrt{2}} & 0\\
        0 &  \frac{1}{\sqrt{2}} &  \frac{1}{\sqrt{2}} & 0\\
         \sqrt{\frac{|g|}{1+|g|}} & 0 & 0 &  \frac{1}{\sqrt{1+|g|}}
    \end{matrix}
    \right),
\end{equation}
where the function sign$(g)$ picks up the sign of $g$. At $g = 1$, we recover the exact toric code. At $g = -1$, the $W$-matrix differs from the toric code $W^{(\text{TC})}$ by a minus sign for one of the vertex configurations. This implies that the wavefunction is related to the toric code ground state by a finite-depth local quantum circuit. However, as we will show below that, the two limits belong to distinct quantum phases due to the presence of physical symmetries. 

The ground state along this path respects an anti-unitary $\mathbb{Z}_2^T$ symmetry generated by the global spin flip composed with complex conjugation $\left(\prod_i X_i\right)K$~\footnote{Note that the wavefunction is real. This is an additional symmetry of the system. We are considering the SET phases protected by the time-reversal symmetry $\left(\prod_i X_i\right)K$, the additional symmetry is helpful for distinguishing the SET order (see Sec.~\ref{sec:qpt}) }. While at $g = 1$, the system is the usual toric code with only an intrinsic $\mathbb{Z}_2$ topological order, at $g = -1$ the TNS describes a ground state with a non-trivial symmetry-enriched topological (SET) order, where the $\mathbb{Z}_2$ topological order is enriched by $\mathbb{Z}_2^T$ symmetry. Along the path, there exists a $\mathbb{Z}_2^T$-symmetric local parent Hamiltonian which remains frustration-free and continuous in $g$ (see Appendix~\ref{sm:parent_H} for details). 

\begin{figure}[t]
    \centering
    \includegraphics{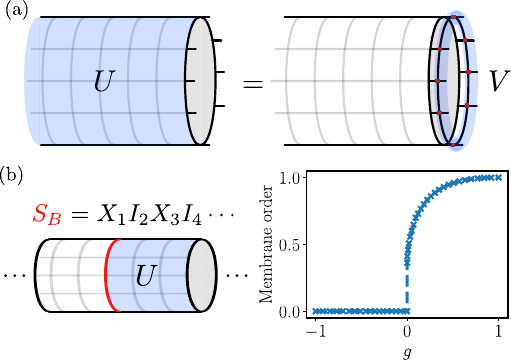}
    \caption{Detecting SET order: (a) Physical symmetry $U$, consisting of Pauli $X$ on all the qubits in the bulk, is mapped to a virtual matrix-product operator (MPO) $V$ on the boundary formed by the uncontracted virtual legs of the tensors. (b) The membrane order $\mathcal{M}$ computed on the minimally entangled ground state corresponding to the anyon $\textbf{e}$ at the thermodynamic limit. The numerical method for extracting the membrane order is described in Ref.~\cite{Haller:2023}. Here $S_B$ is the appropriated string operator inserted at the boundary of the partial action of the physical symmetry. }
    \label{fig:sym}
\end{figure}

Recall that a topologically ordered system has a non-trivial SET order when the anyonic excitations of the system transform non-trivially under the physical symmetries~\cite{Essin:2013,Mesaros:2013,Tarantino:2016,Barkeshli:2019}. The SET order in the ground state can be diagnosed by inspecting how the action of the physical symmetries on the bulk relates to a non-trivial action on the open boundary of the system; the boundary action carries labels that characterize the phases of the system~\cite{Sahinoglu:2014,williamson:2016,Williamson:2017}. To see this explicitly, we consider a cylindrical geometry with open ends. At the boundary, the uncontracted virtual bonds of the TNS form an effective 1D system (see Fig.~\ref{fig:sym}a). Suppose a physical symmetry $U$ is applied to the ground state in the bulk; this physical action is equivalent to a virtual action of a boundary operator $V$. Using the TNS defined by Eq.~\eqref{eq:set_path} on a cylinder with circumference $L\in 4\mathbb{N}$, we have 
\begin{equation}
    U = \left(\prod_i X_i\right)K\to V(g) = \left(\prod_b X_b\right)\text{sign}(g)^{\frac{n(n-1)}{2}},
\end{equation}
where $b$ labels the virtual bonds forming the effective boundary spin-1/2 system and $n = \sum_b (1-Z_b)/2$ is a sum of Pauli $Z$ matrices at each virtual bond ($n$ essentially counts the number of states $\ket{1}$ on the boundary). For $\mathbb{Z}_2^T$ symmetry, a discrete label for the SET phase can be obtained from $V(g)^*V(g) = \text{sign}(g)^{P}$, where $*$ denotes the complex conjugation and $P$ is the parity of the boundary spin system. Different parity $P$ corresponds to distinct anyon labels carried by the minimally entangled states~\cite{Zhang:2012}. In our convention, $P = 0$ labels the trivial quasiparticle while $P = 1$ labels the vertex excitation (i.e. the \textbf{e} anyon) in the toric code. For a given parity, the quantity $V(g)^*V(g)$ is a discrete invariant that labels the elements in the second cohomology group $H^{(2)}(\mathbb{Z}_2^T, \mathbb{Z}_2) = \mathbb{Z}_2$, which classifies the symmetry fractionalization patterns on the anyons in $\mathbb{Z}_2$ topological order under $\mathbb{Z}_2^T$ symmetry (without permutation of anyons)~\cite{Essin:2013,Lu:2016}. Therefore, the symmetry fractionalization patterns on the \textbf{e} anyon are distinct for $g>0$ and $g<0$, the system belongs to different SET phases, with the phase transition occurring at $g = 0$.   

Distinct SET orders cannot be measured via any local order parameters, but they can be distinguished via a non-local membrane order parameter~\cite{Huang:2014, zaletel:2014}, which generalizes the usual string order parameters for 1D symmetry-protected topological (SPT) phases~\cite{Nijs:1989,Kennedy:1992}. To construct a suitable membrane order parameter, we note that the ground-state wavefunction has the additional symmetry of being real. The action of the $\mathbb{Z}_2^T$ symmetry is therefore similar to that of an on-site global spin flip. Consider the system on an infinite cylinder and let $L$ be the circumference, a suitable membrane order parameter is then given by a partial global spin flip $\mathcal{M} = \lim_{L\to\infty} |\bra{\psi} S_B^{(L)} U_{\text{subsys}} S_B^{(R)} \ket{\psi}|^{1/L}$, where $ U_{\text{subsys}} = \prod_{i\in\text{subsys}}X_i$ acts only partially on the bulk and the closed string operator $S_B^{(L,R)} = X_1I_2X_3I_4\cdots X_{L-1}I_L$ around the cylinder are the operators on the left/right boundary of $U_{\text{subsys}}$. The existence of the additional symmetry implies that the membrane order parameter $\mathcal{M}$ vanishes in a non-trivial SET phase protected by $\left(\prod_iX_i\right)K$ due to a superselection rule~\cite{Ppollmann:2012}. In Fig.~\ref{fig:sym}b, we show the membrane order for the minimally entangled state labeling the \textbf{e} anyon. At $g<0$, a non-trivial symmetry fractionalization over the anyon leads to $\mathcal{M} = 0$, while $\mathcal{M}\neq 0$ for $g>0$ corroborates the absence of non-trivial symmetry fractionalization.

\begin{figure}[t]
    \centering
    \includegraphics{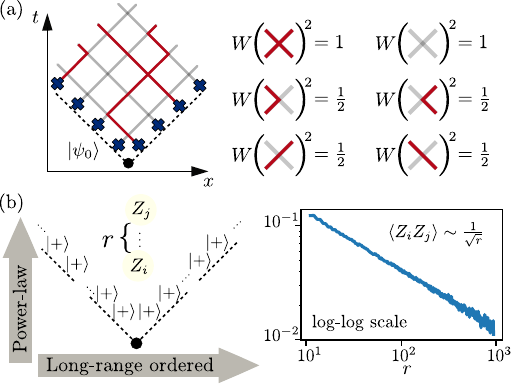}
    \caption{Quantum critical point at $g = 0$: (a) The system in Fig.~\ref{fig:isotns}c is rotated 45 degrees anticlockwise. The isoTNS is contracted from some boundary state $\ket{\psi_0}$ formed by the boundary qubits (dark blue crosses). An example configuration in the wavefunction is shown. The red line denotes $\ket{1}$ and the grey line denotes $\ket{0}$. The particles are located at the endpoints of the red lines and emanate from the boundary (dashed line). They move according to stochastic jumps determined by the $W$-matrix. (b) The correlation along the $t$-direction averaging over $2\times 10^6$ classical trajectories. Here $\ket{+} = (\ket{0}+\ket{1})/\sqrt{2}$. The correlator shows power law decay with power $1/2$.}
    \label{fig:corr}
\end{figure}
\section{Power-Law Correlation at the Quantum Critical Point and Dynamics of 1D Classical Systems}
Along the entire isoTNS path, the classical weight matrix associated with the isoTNS from Eq.~\eqref{eq:set_path} is the same as the weight matrix of the classical eight-vertex model which is exactly solvable~\cite{Baxter:1982zz}.
At the SET transition point $g = 0$, the classical eight-vertex model reduces to a gapless point of the six-vertex model. Interestingly, the corresponding transfer matrix of the six-vertex model has the same spectrum as the Hamiltonian of the 1D ferromagnetic Heisenberg XXX chain (they are solved by the same Bethe Ansatz~\cite{Baxter:1982zz}), exhibiting a spectral gap closing of $O(1/L^2)$. The Heisenberg chain has a long-range order in space and it can support power-law temporal correlation. By analogy, we expect the 2D critical wavefunction to exhibit a similar anisotropy in the correlation functions along the two spatial directions.

In the gapped phases (i.e. $g\neq 0$), the boundary conditions are irrelevant for the bulk properties. However, they become important at the transition point $g = 0$. It is convenient to rotate the lattice in Fig.~\ref{fig:isotns}b anticlockwise by 45 degrees and consider the system on a planar geometry with open boundaries. The isoTNS is then obtained by initializing the tensor network contraction from an arbitrary 1D boundary state $\ket{\psi_0}$ formed by the qubits at the bottom boundary (see Fig.~\ref{fig:corr}a). The resulting critical wavefunction respects a conservation law: The number of lines formed by states $\ket{1}$ is conserved across any horizontal slice, unless they terminate on the boundary. A snapshot of the wavefunction is given in Fig.~\ref{fig:corr}a. This conservation law reveals a direct relation between the quantum critical ground state and the dynamics of an 1D classical system. Suppose the boundary state $\ket{\psi_0}$ is a product state in the Pauli-$Z$ basis and we interpret the endpoint vertex of a string of state $\ket{1}$ as being occupied by a particle. In this picture, the conservation law is simply that of the particle number. As the tensor network is sequentially contracted from the bottom to the top, the particles are moving forward in time and tracing out their worldlines (the $x$- and $t$-direction in Fig.~\ref{fig:corr}a). The Floquet-type dynamics are generated by stochastic jumps with probability given by the matrix elements $|W_{ijmn}|^2$. At a given time $t$, a single particle either moves to the left or to the right along the $x$-direction with an equal probability. If two particles meet at the same vertex, they bounce away from each other. The classical picture suggests that the critical wavefunction is a superposition of all the worldlines of the particles emanating from the boundary.

A few properties of the critical wavefunction follow directly from the classical picture. (i) The critical wavefunction can have a long-range order along the $x$-direction. This long-range order comes from the long-range order in the boundary state $\ket{\psi_0}$ due to the particle-number conservation: The number of states $\ket{1}$ in any row along the $x$-direction is determined by the number of states $\ket{1}$ in $\ket{\psi_0}$. (ii) The critical wavefunction can support power-law correlations along the $t$-direction. To see this, suppose the boundary state $\ket{\psi_0}$ contains only a few particles. The moving particles have a small chance of interacting with each other and the process thus behaves similarly to an independent random walk with diffusive dynamics. More generally, the process is an example of the non-unitary Floquet XXX model~\cite{Pvenicat:2018,friedman:2019,kos:2021}. We verify the power-law correlation by computing the spin-spin correlation $\langle Z_iZ_j\rangle$ between sites $i,j$ in the bulk along the $t$-direction for the boundary condition $\ket{\psi_0} = \ket{++\cdots +}$, where $\ket{+} = (\ket{0} + \ket{1})/\sqrt{2}$. This boundary condition is an equal-weight superposition of all the possible classical initial conditions. The resulting correlation corresponds to the typical correlation function over this ensemble. The result is shown in Fig.~\ref{fig:corr}b, a clear power-law decay as $1/\sqrt{r}$ can be seen, which is consistent with the diffusive behaviour. In Appendix~\ref{sm:aniso_corr}, we show that the connected correlator along a direction between the $t$- and the $x$-direction falls off exponentially.

\begin{figure}[t!]
    \centering
    \includegraphics{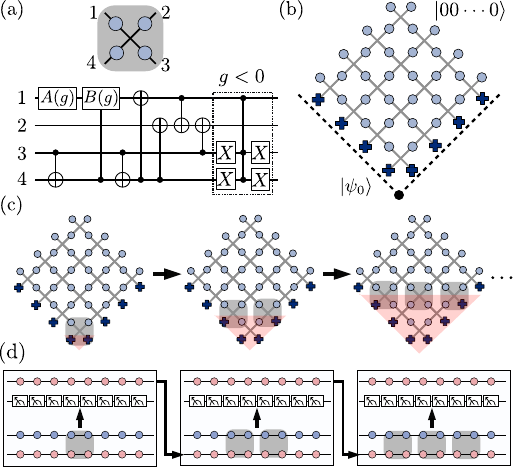}
    \caption{The explicit quantum circuit and the holographic quantum algorithm. (a) A parametrized 4-qubit gate generating the isoTNS. The gate is decomposed into a single-qubt gate, two-qubit CNOT gates and a control-rotation gate. An additional three-qubit gate is included for $g<0$. Here, $A(g)$ and $B(g)$ are single-qubit gates such that $A(g) \ket{0}= \sqrt{1/(1+|g|)}\ket{0} + \sqrt{|g|/(1+|g|)}\ket{1}$ and $B(g) = HA(g)^{\dag}$, where $H$ is the Hadamard gate. The three-qubit gate is a CCZ gate such that $\text{CCZ}\ket{111} = -\ket{111}$ and it acts trivially otherwise. (b) Initialization. The physical qubits (blue circles) are located at the edges of a square lattice. The system is initialized as $\ket{\psi_0}\otimes\ket{00\cdots 0}$, where $\ket{\psi_0}$ is the boundary state. (c) Generating the isoTNS. The gate in (a) is applied diagonally in parallel. The red shade marks the spread of the entanglement. (d) Holographic simulation of the isoTNS. At each step, we implement the sequential circuit on both the physical qubit array (red) encoding the boundary of the partially realized wavefunction and the the unentangled ancillary qubit array initialized as $\ket{00\cdots 0}$ (blue), then we perform projective measurements in the eigenbasis of each single-site operator $O_k$. The post-measurement qubits are reset and reused as the ancillary qubits in the next step.}
    \label{fig:circuit}
\end{figure}

\section{Quantum simulation of 2D quantum phase transitions represented by isoTNS}

The isoTNS path above can be directly realized experimentally on a 2D array of qubits using an efficient linear-depth quantum circuit.
In this section, we present the explicit quantum circuit for the path considered in the previous section. Furthermore, we show that the isoTNS path can also be simulated with a hardware-efficient holographic quantum algorithm. We further show how to characterize the SET phase transition experimentally. This provides a minimal proposal for a proof-of-principle experiment for the simulation of a 2D topological quantum phase transition on programmable quantum hardware.  
\subsection{Explicit quantum circuit representation}\label{sec:circ}

The 2D quantum circuit representation for the isoTNS path can be easily found when the system is defined with open boundaries~\footnote{For systems with a periodic boundary condition, there is no clear way to initialize the contraction of the isoTNS}. The isoTNS tensor is mapped to a 4-qubit quantum gate acting on each vertex, as shown in Fig.~\ref{fig:circuit}a. The system is initialized as  $\ket{\psi_0}\otimes\ket{00\cdots 0}$, where $\ket{\psi_0}$ is any 1D quantum state formed by the boundary qubits (see Fig.~\ref{fig:circuit}b). The state is generated by sequentially applying the 4-qubit gate to each vertex along the diagonal direction, as depicted in Fig.~\ref{fig:circuit}c. The depth of the circuit thus scales as $O(L)$ as long as the boundary state $\ket{\psi_0}$ can be prepared with an $O(L)$-depth circuit. Using the exact quantum circuit, the ground states along the entire path can be realized without any post-selection.

\subsection{Holographic quantum simulation of isoTNS}\label{sec:holo}

It has been shown that 1D MPS of arbitrary length can be simulated on a quantum simulator with a constant number of qubits by performing the contraction of the MPS temporally. This provides a hardware-efficient way to simulate MPS on quantum computers (see e.g. Ref.~\cite{foss:2021}). Here we show that the same idea can be used to simulate the 2D isoTNS requiring only an 1D array of qubits. 

The sequential quantum circuit for preparing the isoTNS (Fig.~\ref{fig:circuit}c) has a simple causal structure. Following the direction of the sequential circuit, the measurements of any local operators supported at the bulk of the wavefunction are unaffected by the quantum gates at the far side along the direction of the sequential circuit. Consequently, the measurement of a tensor product of single-site Hermitian operators $\langle O_1 O_2\cdots  O_N\rangle $ is equivalent to a sequential measurement protocol as follows:
\begin{enumerate}
    \item Following the orientation in Fig.~\ref{fig:circuit}c, we apply the first step of the sequential circuit and measure the single-site observable $O_k$ at the first row of qubits. Each measurement is performed in the basis of $O_k$ for each site $k$.
    \item Proceed to the next step of the sequential circuit and increment to the next row of qubits for the measurement, repeat 1.
\end{enumerate}
The steps are summarized in Fig.~\ref{fig:circuit}d. Note that a key feature of the sequential measurement is that at each step, the measurements are not performed on the full isoTNS, but on the boundary of the partially prepared isoTNS. Due to the causal structure of the isoTNS and commutativity of the measurements on each row, the sampling based on the sequential algorithm is the same as directly sampling $\langle O_1 O_2\cdots  O_N\rangle $ on the full isoTNS wavefunction. More generally, a tensor product of local observables can be efficiently sampled with the same strategy by extending the number of gates included in the lightcone of the isoTNS.


After each step of the sequential measurement, the post-measurement qubit at site $k$ collapses into some eigenstate of $O_k$ and becomes completely disentangled from the system. The disentangled qubits can then be reset back to states $\ket{0}$ and reused for the next step of the sequential measurement, as depicted in Fig.~\ref{fig:circuit}d. Then the whole measurement process can be performed with only two 1D arrays of qubits (which can be implemented by the even and odd sites of a 1D array of qubits). Since at each step, the quantum system stores the 1D boundary state of the partially prepared isoTNS conditioned on the measurement outcomes, this is a generalization of the holographic quantum algorithm of 1D MPS to 2D isoTNS. 

The holographic quantum algorithm makes it easy to measure general observables. The entanglement entropy can also be accessed via the technique of randomized measurements (which is based on sampling bitstrings on a random basis)~\cite{ Elben2018, brydges:2019,Satzinger2021}. A signature of the phase transition can in principle be captured by observing a diverging correlation length extracted from the measurement of some two-point correlators. However, the measurement of the membrane order parameter used in Sec.~\ref{sec:qpt}, which distinguishes the SET phases, relies on the preparation of minimally entangled states in a cylindrical geometry and is therefore experimentally challenging to achieve. In the next section, we discuss how the SET order parameter can be probed without creating the minimally entangled states and can be measured using the holographic quantum algorithm for the isoTNS.

As a side remark, the sequential measurement also provides a classical numerical method for sampling 2D isoTNS. Since an MPS with a finite bond dimension can be efficiently sampled, the sequential measurement can be simulated by representing the 1D boundary state as an MPS (conditioned on the sequential measurement outcomes from the previous steps). If the 1D boundary state can be efficiently approximated by an MPS with a low bond dimension, then we expect that the sampling of the isoTNS is classically simulable, otherwise (i.e. when the MPS has a volume-law entanglement) the classical simulation is hard. This is in spirit similar to the classical numerical method proposed for sampling 2D shallow quantum circuits~\cite{napp:2022}.

\subsection{Probing the SET phase transition}
A defining feature of the non-trivial SET phase is that the symmetry action fractionalizes non-trivially around the anyon. A straightforward order parameter for the SET phases is obtained by measuring the membrane order parameter in Sec.~\ref{sec:qpt} around a single anyon in the bulk. For a real wavefunction, the membrane order parameter will vanish if there is a non-trivial symmetry fractionalization of the time-reversal symmetry and remain non-zero otherwise. 

The difficulty of this approach lies in the creation of a single anyon in the bulk. While this is easy at the two fixed points of the toric code ($g = \pm 1$), where the anyon creation operators are some string-like operators that are exactly known and are simple to realize, efficient ways to create and manipulate the anyons away from the fixed-point limits are not a priori known. 

\begin{figure}
    \centering
    \includegraphics{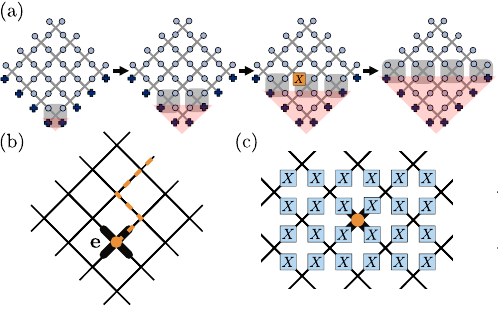}
    \caption{An illustration of a defect (\textbf{e} anyon) insertion into the isoTNS wavefunction. (a) A defect can be inserted by applying an additional Pauli $X$ flip to an output qubit after a step of the sequential circuit. The plots show the first four steps of the state preparation, a defect is inserted after step 3. (b) The resulting wavefunction corresponds to the ground state with an \textbf{e} anyon at the defected vertex. This is equivalent to generating an \textbf{e} anyon by applying some (dressed) string operator connecting the defect and the open boundary. (c) The membrane order parameter consists of a product of Pauli $X$ on adjacent non-overlapping plaquettes surrounding the defect.}
    \label{fig:defect}
\end{figure}
The use of the isoTNS provides a solution to the problem. To create an anyon, we simply need to replace one of the local tensors in the isoTNS by a single ``defect'' tensor. The defect tensor is still required to satisfy the isometry condition listed in Eq.~\eqref{eq:isometry}, but unlike the original local tensor where states around each vertex sum to an even parity (see Eq.~\eqref{eq:set_path}), the defect tensor only allows for those states that are summed to an odd parity to meet at the vertex. The rest of the tensors in the isoTNS stay the same. As a result, the isoTNS wavefunction with a single local defect tensor is orthogonal to the ground state. Physically, the isoTNS with a defect corresponds to the ground state with a single vertex excitation (or an \textbf{e} anyon in the toric code, potentially dressed by some local pairs of plaquette excitations, or \textbf{m} anyons) at the defect vertex. Since the defected isoTNS is still an isoTNS, it can be efficiently generated and simulated via the sequential protocols. An excitation above the ground state is created without the use of explicit particle-creation operators.

At a quantum circuit level, inserting a defect tensor is easy to achieve. In Fig.~\ref{fig:defect}a, we show one way of inserting defect by simply flipping an output qubit after one of the sequential steps. The sequential steps afterwards stay the same as before. Note that on a compact manifold, we expect all anyonic excitations above the $\mathbb{Z}_2$ topologically ordered ground state to occur in pairs. Here a single anyon in the bulk is made possible by the open boundary condition, with one anyon in the bulk and the other removed to the boundary of the system (see Fig.~\ref{fig:defect}b).

Once we insert an \textbf{e} anyon, a membrane order parameter $\mathcal{M}$ can be measured,
\begin{equation}
    \mathcal{M} = \lim_{L\to\infty}|\langle \psi|\prod_{i\in P_e}X_i|\psi\rangle|^{1/L},
\end{equation}
where $\ket{\psi}$ is the wavefunction with an inserted defect and $P_e$ denotes a set of sites on adjacent non-overlapping plaquettes around the anyon (as shown in Fig.~\ref{fig:defect}c), and $L$ is the perimeter of the supported region of this membrane. Similar to the membrane order in Sec.~\ref{sec:qpt}, for a real wavefunction, the membrane order vanishes at the non-trivial SET phase protected by $(\prod_iX_i)K$. In a trivial SET phase, $\mathcal{M}$ is generically non-zero. Since $\mathcal{M}$ can be measured from a product of Pauli $X$, this can be directly measured from the holographic simulation of the isoTNS path.

\section{Discussion and outlook}
To summarize, we have proposed a framework to design efficient quantum circuits for the quantum simulation of 2D topological quantum phase transitions using the tool of isoTNS. We achieved this by restricting to a subclass of isoTNS via the plumbing construction, which connects the resulting isoTNS wavefunctions to classical partition foundations. We illustrated the method by constructing a simple example of an $D = 2$ isoTNS path between distinct SET phases. Furthermore, we proposed a minimal experimental protocol for the simulation and the detection of the SET phase transition on a programmable quantum computer based on the holographic quantum simulation of 2D isoTNS. The proposed framework highlights the application of isoTNS in the context of quantum simulation.

While we have illustrated the design of an efficient circuit based on a specific isoTNS path interpolating between distinct SET phases, the method can be applied more generally. A large class of topologically ordered states admits exact isoTNS representation where the virtual legs and the physical legs can be related by the plumbing tensor (e.g. string-net states~\cite{soejima:2020}). One can start from these fixed-point isoTNS and apply the plumbing method to prepare other ground states away from the fixed points and potentially cross a quantum phase transition to reach a distinct phase of matter. For example, we show in Appendix~\ref{sec:sm_tcds} that by adapting the plumbing method to the domain walls of tensor-network virtual legs (the double-line TNS~\cite{Gu:2008}), an isoTNS path with bond dimension $D = 4$ can be constructed between the toric code and the double-semion ground state with distinct topological order~\cite{Freedman:2004,levin:2005}, crossing the same critical point as the SET case. The isoTNS path is analogous to the $D = 4$ non-isometric TNS path studied in Ref.~\cite{xu:2018}, which crosses a critical point with a different six-vertex criticality. 

Another example is the tensor-network solvable path noted in Ref.~\cite{Tantivasadakarn:2021}. It can be verified that the path is in fact an isoTNS path that can be constructed with the plumbing method. This path interpolates between two 1-form SPT phases with the exact toric code ground state being the critical wavefunction. An exciting future direction is to explore the possibility of connecting generic gapped quantum phases via continuous isoTNS paths.
It is worth noting that our construction can also be related to the concept of strange correlators~\cite{you:2014,vanhove:2018} by taking the overlap between the wavefunction and an appropriately chosen product state. The overlap can be interpreted as a partition function defined with complex Boltzmann weights given by the entries of the $W$-matrix. The properties of the strange correlators for the isoTNS path might be helpful for a systematic understanding of connecting gapped phases with isoTNS.

It has been known that typical correlations in isoTNS decay exponentially~\cite{haag:2023}. An open question remained, however, whether isoTNS can also support power-law correlations. In this work we present a concrete example, which demonstrates that this is indeed possible. More intriguingly, we find that the power-law decay of correlations in this example originates from a direct connection between the worldlines in 1D stochastic dynamics and 2D isoTNS ground states. Whether this connection can lead to a systematic construction of more general isoTNS remains an open question. Incidentally, the efficient physical realization of 2D topological quantum phase transitions also presents a valuable resource for performing and benchmarking algorithms for quantum phase recognition in dimensions higher than one~\cite{huang:2022,liu:2023}.

\section{Acknowledgement}
We thank Michael Knap, Robijn Vanhove, Wen-Tao Xu, Xie-Hang Yu and Toma\ifmmode \check{z}\else \v{z}\fi{} Prosen for inspiring discussions.
Y.-J.L. and F.P. acknowledge support from the Deutsche Forschungsgemeinschaft (DFG, German Research Foundation) under Germany’s Excellence Strategy--EXC--2111--390814868  and the TRR 360 – 492547816, the European
Research Council (ERC) under the European Union’s Horizon
2020 research and innovation programme (grant agreement No. 771537), as well as the Munich Quantum Valley, which is supported by the Bavarian state government with funds from the Hightech Agenda Bayern Plus.

{\par\textbf{Data and materials availability:}} Data analysis and simulation codes are available on Zenodo upon reasonable request~\cite{zenodo}.

\bibliography{isotns}

\appendix

\section{The plumbing construction in 1D MPS}\label{sm:mps_plumb}
The plumbing method can be applied to the 1D system and we recover some of the familiar examples of quantum phase transitions in 1D matrix-product states (MPS)~\cite{wolf:2006}. MPS are an ansatz class where the coefficients of a full 1D $N$-qubit state $\ket{\psi}$ are decomposed into products of matrices. Explicitly, for a system with open boundaries and translational invariance in the bulk
\begin{equation}\label{sm:eq:mps}
    \begin{split}
    \ket{\psi} = \sum\limits_{\{\sigma\}}&A^{\sigma_1}B^{\sigma_2}B^{\sigma_3}\dots B^{\sigma_{n-1}}C^{\sigma_n}\ket{\sigma_1\sigma_2\sigma_3\dots\sigma_n},
    \end{split}
\end{equation}
where the $\sigma_k\in\{0,1,\cdots,d\}$ indices are the physical indices and $B^{\sigma}$ are $\chi\times\chi$ matrices with $\chi$ being the bond dimension of the MPS. The boundary tensors $A^\sigma$ and $C^{\sigma}$ are $1\times \chi$ and $\chi\times 1$ matrices, respectively. Analogous to the main text, we employ the plumbing structure and impose the isometry condition. Namely, we require
\begin{equation}
    B^{\sigma}_{ij} = \sum_{j'}\delta^{\sigma}_{ij'} W_{j'j}, 
\end{equation}
where the $\delta$-tensor takes value 1 when all indices are the same and zero otherwise. Furthermore
\begin{equation}
    \sum_{\sigma,j'} \left(B^{\sigma}_{ij'}\right)^* B^{\sigma}_{jj'} = \delta_{ij}.
\end{equation}
This isometry condition is satisfied if and only if
\begin{equation}
    \sum_{j} |W_{ij}|^2 = 1,\ \forall i.
\end{equation}
Now suppose we take $C^\sigma_i = \delta^{\sigma}_i$ and $A^\sigma$ is some boundary condition that can be chosen at will. The MPS in Eq.~\eqref{sm:eq:mps} is essentially a canonical form of the MPS~\cite{Perez-Garcia:2006}, which is equivalent to a sequential quantum circuit~\cite{schon:2005}. 

As an example, consider a $W$-matrix path with $d = \chi = 2$, for $g\in[-1,1]$
\begin{equation}
    W(g) = 
\begin{pNiceMatrix}[first-row,last-col]
 \ket{0}   & \ket{1} &      \\
    \frac{1}{\sqrt{1+|g|}}  & \text{sign}(g)\sqrt{\frac{|g|}{1+|g|}}   & \ \ket{0} \\
    \sqrt{\frac{|g|}{1+|g|}}   &   \frac{1}{\sqrt{1+|g|}}    & \ \ket{1} 
\end{pNiceMatrix}.
\end{equation}
At $g = 1$, the wavefunction $\ket{\psi(g)}$ is a simple product state $\ket{++\cdots +}$ in the bulk. At $g = -1$, the state is a cluster state with $Z_iX_{i+1}Z_{i+2}\ket{\psi(-1)} = -1$ for all $i$ in the bulk. At these two limits, the 1D system belongs to distinct 1D SPT phases protected by anti-unitary $\mathbb{Z}_2^T$ symmetry generated by $\left(\prod_i X_i\right)K$, where $K$ is the complex conjugation. It can be verified that this is a symmetry for the state for all $g$.

A quantum phase transition happens at $g = 0$, where the wavefunction has a long-range order and becomes the Greenberger–Horne–Zeilinger (GHZ) state $\ket{\psi(0)} = (\ket{00\cdots 0} +\ket{11\cdots 1})/\sqrt{2}$ with the boundary tensor chosen as $A^{\sigma}_i = \delta^{\sigma}_i$. In fact, the MPS along this path is exactly the canonical form of the $\chi = 2$ MPS considered in Ref.~\cite{wolf:2006}, with a parent Hamiltonian
\begin{equation}
     H(g) = g_{zxz}\sum_i Z_{i-1}X_iZ_{i+1} - g_{zz}\sum_i Z_iZ_{i+1} - g_x\sum_i X_i,
\end{equation}
where $g_{zxz} = (1-g)^2, g_x = (1+g)^2$ and $g_{zz} = 2(1-g^2)$. This canonical form has been utilized to construct an efficient quantum circuit for the physical realization of this phase transition on a digital quantum computer~\cite{smith:2022}.

\section{General isoTNS and their quantum circuit representation}\label{sm:isotns_circ}
\begin{figure}[t!]
    \centering
    \includegraphics{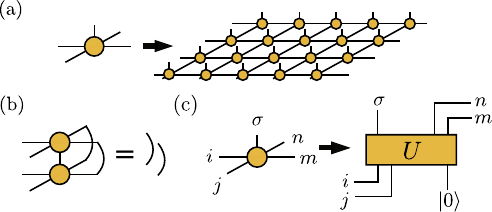}
    \caption{The schematics for 2D TNS (a) The wavefunction is contracted from local tensors (yellow circles) with a finite bond dimension. (b) and (c) illustrate the isometric tensor networks. (b) shows the isometric condition on the local tensors. (c) An isoTNS can be sequentially generated by applying a unitary $U$ sequentially to a product state )with ancillary qudits). This unitary is obtained by lifting the isometric local tensors to a full unitary.}
    \label{fig:2dtns_intro}
\end{figure}

In this section, we briefly review the concept of isoTNS and their quantum circuit representation.
A 2D tensor-network state (TNS) (also known as Projected Entangled Pair States (PEPS)~\cite{verstraete2004renormalizationalgorithmsquantummanybody}) is obtained by the contraction of local tensors. Each local tensor $T^{\sigma}_{ijmn}$ consists of four virtual legs with bond dimension $D$ and one physical leg with local physical dimension $d$. The wavefunction for an $N$-site state is
\begin{equation}
    \ket{\psi} = \sum\limits_{\{\sigma_s\}} \text{tTr}\left(T^{[1]\sigma_1}T^{[2]\sigma_2}\cdots T^{[N]\sigma_N}\right)\ket{\sigma_1,\sigma_2,\cdots,\sigma_N},
\end{equation}
where tTr denotes the 2D tensor contraction of the nearest neighbour virtual legs (see Fig.~\ref{fig:2dtns_intro}a). Any 2D wavefunctions can be written as 2D TNS with bond dimension exponentially large in the system size. It is believed that the ground states of local gapped 2D Hamiltonian can be well approximated by 2D TNS with a low bond dimension. For an extensive review of 2D TNS, see Ref.~\cite{cirac:2021}. Note that compared to the tensor in Fig.~\ref{fig:isotns} with two physical legs, each tensor in Fig.~\ref{fig:2dtns_intro}a has a single physical leg with a larger local dimension. The latter contains the former as a special case and is a more general expression for the 2D TNS.

\begin{figure}[t!]
    \centering
    \includegraphics[scale = 0.6]{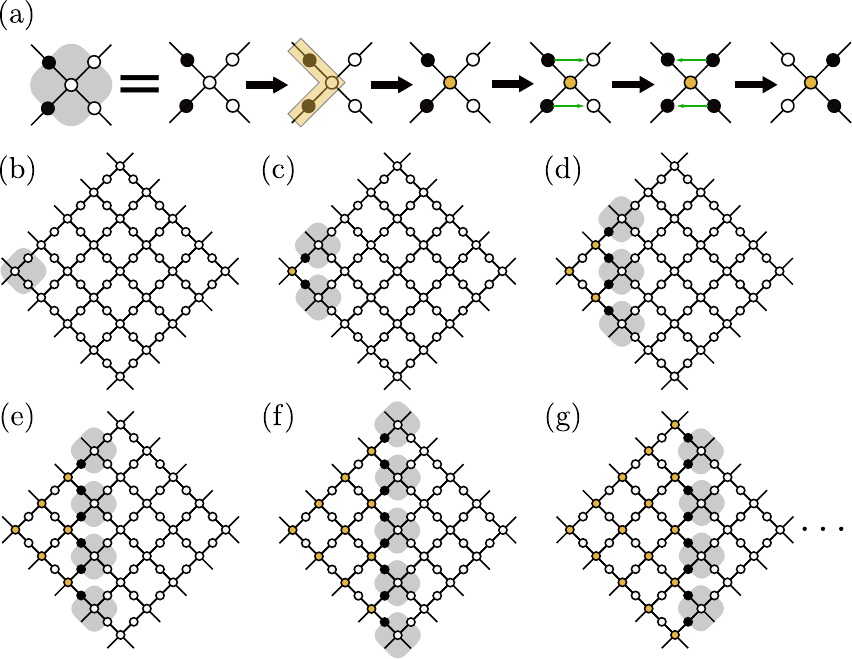}
    \caption{A sequential quantum circuit for general isoTNS. The white circles are disentangled qudits in state $\ket{0}$. The black circles are the ancillary qudits storing the uncontracted virtual legs (legs $i.j$ in Fig.~\ref{fig:2dtns_intro}c). The yellow circles are the entangled physical qudits of the prepared wavefunction. (a) At each step, a five-qudit gate is applied around each vertex. The gate first applies the lifted unitary (yellow block) in Fig.~\ref{fig:2dtns_intro}c, where the black qudit states on the links are taken as the input legs $i.j$. After the lifted unitary, The uncontracted virtual legs (legs $m,n$ in Fig.~\ref{fig:2dtns_intro}c) are still stored in the black qubit states. Next, these ancillary states are swapped to the ancillary qudits on the right. (b)-(g) The isoTNS is generated by applying the unitary in (a) along each column of vertices sequentially. At the end of the circuit, the ancillary qubits are all in disentangled states $\ket{0}$ and the physical qudits store the isoTNS.}
    \label{fig:iso_intro_circ}
\end{figure}

Generic 2D TNS do not have a well-defined canonical form and are hard to prepare on a quantum computer~\cite{schuch:2007}. A subclass of general 2D TNS, the isometric tensor-network states (isoTNS)~\cite{zaletel:2020}, shares the same property as 1D MPS that they can be efficiently prepared with a sequential quantum circuit with depth $O(L)$, where $L$ is the linear size of the system. Surprisingly, despite being a subclass of general 2D TNS, isoTNS are expressive and are known to be able to exactly represent a large class of gapped many-body ground states. For this reason, isoTNS serve as a valuable tool not only in the numerical method, but they can also be utilized to design efficient quantum circuits for interesting classes of many-body ground states.

IsoTNS are 2D tensor-network states where the local tensors additionally satisfy an isometric condition, as depicted in Fig.~\ref{fig:2dtns_intro}b. Explicitly, we have
\begin{equation}
    \sum_{m,n,\sigma} \left(T^{\sigma}_{ijmn}\right)^* T^{\sigma}_{i'j'mn} = \delta_{ii'}\delta_{jj'},
\end{equation}
where $\delta_{ii'}$ is the kronecker delta and takes 1 if $i = i'$ and zero otherwise.
This isometric condition implies that each local tensor can be directly lifted to a local unitary which can be used to efficiently generate the isoTNS (see Fig.~\ref{fig:2dtns_intro}c).

We illustrate how a sequential quantum circuit for general 2D isoTNS can be constructed. Suppose we want to prepare a finite 2D isoTNS with open boundary condition. We begin with a square lattice with physical qudits (a $d$-level quantum system) at each vertex and ancillary qudits (a $D$-level quantum system, where $D$ is the bond dimension of the TNS) on each link. We initialize the sequential circuit from the boundary closer to the input virtual legs of the lifted local unitary (left-pointing legs of Fig.~\ref{fig:2dtns_intro}c). The qudits on this boundary are initialized according to the chosen boundary condition. The rest of the qudits are all initialized in state $\ket{0}$. The isoTNS is generated by applying the vertex unitary in Fig.~\ref{fig:iso_intro_circ}a sequentially to each column of the vertices (see Fig.~\ref{fig:iso_intro_circ}b-g). The application of the gates across each column is effectively contracting the isoTNS tensors along the column.  At each step, the ancillary qudits store the uncontracted virtual legs of the local tensors. They are disentangled back to states $\ket{0}$ at the end of the circuit.

The use of the ancillary qudits simplifies the structure of the circuit but they are not necessary to achieve the preparation. The ancillary states can also be stored in the disentangled physical qudits that are not yet reached by the sequential circuit. With a careful design of the gates, the resulting sequential circuit without ancillary qudits also consists of $O(L)$ layers of local quantum gates~\cite{wei:2022}.

In the main text, we construct the isoTNS via the plumbing method, which locks the physical legs with the virtual legs on each tensor. As a result, the uncontracted virtual legs can be stored as the physical qubits and no ancillary qubits are needed. This simplification leads to the simple quantum circuit presented in Sec.~\ref{sec:circ} for the isoTNS path.

\section{Symmetry condition on local tensors for TNS wavefunction}\label{sm:sym}
As mentioned in the main text, compared to the quantum-circuit representation, the tensor-network representation provides a convenient framework for handling the symmetries in the represented states. The ability to preserve symmetries are particularly important when using variational states to represent quantum phase transitions in the presence of topological order and some physical symmetries.

\textbf{Instability of topological order in 2D TNS.} Suppose we have a topologically ordered 2D TNS contracted from the local tensor $T^{\sigma}_{ijmn}$. Now if we slightly vary the local tensor by $T^{\sigma}_{ijmn} + \epsilon M^{\sigma}_{ijmn}$, where $M^{\sigma}_{ijmn}$ is some perturbation tensor with bounded entries and $\epsilon$ is a small number. It has been shown that for generic $M^{\sigma}_{ijmn}$, the topological order of the 2D TNS is unstable for any $\epsilon\neq 0$! Note that this is not contradicting the general understanding that topologically ordered systems are stable under perturbation. The instability could arise because the perturbation is done on the virtual level of the tensor networks, rather than on the physical degrees of freedom. A perturbation on the local tensor can generically correspond to non-local perturbation on the physical level, resulting in a breakdown of topological order~\cite{shukla:2018}. 

A remedy for the instability is to identify and preserve the same virtual symmetries of the local tensor as in the unperturbed case during the variation. On an abstract level, imposing the correct virtual symmetries restricts the local tensor to the \textit{MPO-injective} subspace such that local variation of the tensors translates to local physical perturbation. The MPO-injective subspace encodes the key data for classification of topologically ordered tensor-network states~\cite{williamson:2016,Williamson:2017}. For the simple example of $\mathbb{Z}_2$ topologically ordered tensor network considered in the main text, the virtual symmetry is simply the closed-loop constraint generated by a product of Pauli $Z$ operators at each of the four virtual legs of a local tensor $T$. We therefore enforce this virtual symmetry along the entire path of Eq.~\eqref{eq:set_path} to maintain the perturbative stability of the topological order. For more general topological order, the virtual symmetries are closely related to the topological data of the topological order and can be systematically constructed for a large class of fixed-point states~\cite{williamson:2016,Williamson:2017}.

\textbf{Sufficient local condition for symmetric wavefunctions.} In order to represent phase transitions between quantum phases protected by symmetries, it is important that the TNS preserves the same symmetry as we vary the parameters. However, symmetry is a global condition and therefore one would expect that this is generally hard to check. Tensor-network representation provides a powerful way for checking the symmetry of the state. This is particularly nice in 1D MPS, where there is a direct correspondence of all on-site symmetries of the state and some condition of the local tensor of the MPS~\cite{Perez-Garcia:2006}. While in 2D TNS, this correspondence does not exist but a sufficient local condition for an on-site symmetry is given by a ``pulling through'' condition~\cite{cirac:2021}   
\begin{equation}
    \includegraphics[scale = 1.1]{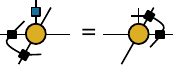}.
\end{equation}
More precisely, there exists a matrix-product operator (MPO) (black boxes with bonds) such that it can be pulled through the tensor acted on by the on-site symmetry operator (blue block). If such MPO exists, then the TNS wavefunction is symmetric under the on-site symmetry. In particular, in the case when the bond dimension of the MPO is 1, this condition reduces to the simple form
\begin{equation}
    \includegraphics[scale = 1.1]{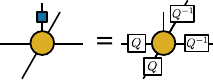},
\end{equation}
where $Q$ is some invertible $D\times D$ matrix, where $D$ is the bond dimension of the tensor.

For the example in the main text, we want to check the symmetry $(\prod_i X_i)K$. By knowing the wavefunction is real, we only need to check the on-site symmetry $\prod_iX_i$ is satisfied. It is easy to verify that for the path Eq.~\eqref{eq:set_path}, this local condition is satisfied by picking $Q$ to be the Pauli $X$ operator for $g>0$ and $Q = XS$ for $g<0$ (both orders of multiplying $X$ and $S$ give a valid $Q$), where $S$ is a $2\times 2$ matrix with matrix elements $(S)_{10} = (S)_{01} = 0$, $(S)_{00} = 1$ and $(S)_{11} = i$.

Besides the analytical approach, for a finite system (e.g. on a long cylinder) the symmetry can also be checked numerically using the approach discussed in Ref.~\cite{Ppollmann:2012}. Explicitly, a modified global transfer operator for the tensor-network wavefunction can be constructed from the overlap of local tensors sandwiching the on-site symmetry operators $\mathcal{T} = \sum_{\sigma,\rho}O_{\sigma\rho}(T^{\sigma}_{ijmn})^*T^{\rho}_{i'j'm'n'}$, where $O_{\sigma\rho}$ is the matrix representation for the on-site symmetry operator. An example of a global transfer operator from 3 local $\mathcal{T}$ around a cylinder is 
\begin{equation}
    \includegraphics[scale = 1]{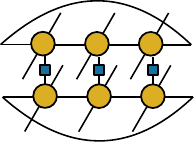}.
\end{equation}
If the 2D TNS represents a (normalized) symmetric wavefunction on an infinitely long cylinder, then the largest eigenvalue $\eta$ of the global transfer operator should satisfy $|\eta| = 1$. If the wavefunction is not symmetric under the symmetry operator, then $|\eta|<1$, indicating a vanishing expectation value of the symmetry operator with respect to the wavefunction as the system size increases. 

\section{Anisotropic correlation at the quantum critical point}\label{sm:aniso_corr}
\begin{figure}
    \centering
    \includegraphics{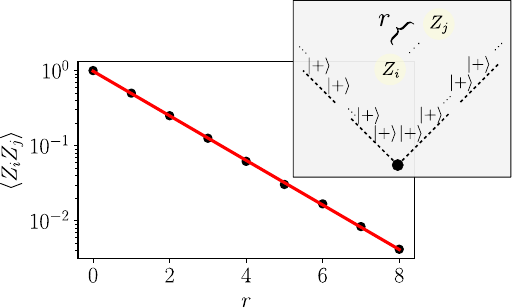}
    \caption{The two-point correlation function along the 45-degree direction exhibits exponential decay (note the log scale of the vertical axis). The black dots are the results from Monte Carlo sampling over $1.5\times 10^6$ samples on a system of $21\times 21$ plaquettes. The red line is a linear fit with a gradient around -0.68.}
    \label{sm:fig:corrdiag}
\end{figure}
In the main text, we show that at the critical point with a boundary state $\ket{\psi_0} = \ket{++\cdots +}$, the two-point correlation along a particular spatial dimension decays algebraically. Along other spatial directions, we expect that the correlation decays exponentially, similarly to the spatial-temporal correlation for a single random walker in the 1D classical dynamics.

We note that along the $x$-direction in Fig.~\ref{fig:corr}a, the spin-spin correlation $\langle Z_iZ_j\rangle$ vanishes due to the disordered boundary state. Furthermore, suppose we now measure the correlation function along the 45-degree direction (see Fig.~\ref{sm:fig:corrdiag}), which decays exponentially. More generally, when $\langle Z_iZ_j\rangle$ is measured along a direction different from the $t$-direction in Fig.\ref{fig:corr}a, we expect that the correlation decays exponentially with an angle-dependent correlation length. 

\section{Quantum-classical correspondence}\label{sm:qc_map}
In this section, we describe the connection between the plumbing construction and the quantum-classical correspondence in tensor networks. We note that the squared norm of the TNS wavefunction represented by the plumbed local tensor $T^{\sigma\rho}_{ijmn}$ can be expressed as the product of the transfer operators $\langle\psi|\psi\rangle = \Tr(M^L)$ (assuming periodic boundary condition on an $L\times L$ system). The transfer operator $M = R_1R_2\cdots R_L$ is the product of the weight matrix $R$ along each row, where $R_{ijmn} = |W_{ijmn}|^2$ and $W$ is the $W$-matrix used in the plumbing procedure. One way to contract the transfer operator can be visualized as
\begin{equation}
    \includegraphics[scale = 1]{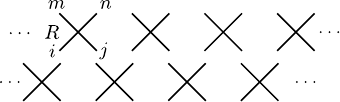}.
\end{equation}
Indeed, one can also rotate each weight matrix $R$ by 45 degrees and contract.
This is exactly the same as contracting a 2D classical partition function.
As an example, consider an $L\times L$ classical spin system on a square lattice with nearest-neighbour two-body interaction $H = \sum_{\langle i,j\rangle} h(\sigma_i,\sigma_j)$. The partition function of the system $\mathcal{Z} = \Tr(e^{-\beta H}) = \Tr(M^L)$ can be conveniently expressed using the transfer matrix $M = R_1 R_2 \cdots R_L$, where $(R_k)_{ijmn} = \sum_{\sigma}\exp\left[-\beta \left(h(i,\sigma)+h(j,\sigma)+h(\sigma,m)+h(\sigma,n)\right)\right]$ is the local weight matrix that carries the statistical weights associated with the coupling between the $k$-the spin along each row and its neighbours. The 2D classical model can thus be mapped to a 2D quantum model using the plumbing construction.



\section{A parent Hamiltonian for the isoTNS path between SET phases}\label{sm:parent_H}
\begin{figure}
    \centering
    \includegraphics{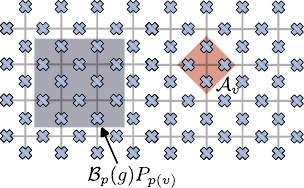}
    \caption{A local parent Hamiltonian for the isoTNS ground states in the main text. The Hamiltonian is frustration-free and consists of projectors around each plaquette (the shaded 12 qubits) and each vertex (shaded 4 qubits). }
    \label{sm:fig:parenth}
\end{figure}
In this section, we derive the parent Hamiltonian for the isoTNS path presented in the main text. Let us consider the toric code described by the Hamiltonian defined on the same square lattice as in the main text,
\begin{equation}\label{sm:eq:htc}
    H_{\text{TC}} = \sum_v \mathcal{A}_v +\sum_p \mathcal{B}_p,
\end{equation}
where $\mathcal{A}_v = (1-\prod_{i\in v} Z_i)/2$ and $\mathcal{B}_p = (1-\prod_{i\in p}X_i)/2$ are projectors made from products of Pauli operators around each vertex $v$ and each plaquette $p$. The vertex and the plaquette operators commute with each other and the ground state $\ket{\text{TC}}$ satisfies $\mathcal{A}_v\ket{\text{TC}} = \mathcal{B}_p\ket{\text{TC}} = 0$ for all $v,p$. For $g> 0$, the isoTNS path between the SET phases can be obtained from the toric code ground state by an imaginary time evolution
\begin{equation}\label{sm:eq:imag}
    \ket{\Psi(g)} = \mathcal{T}(\beta_1, \beta_2)\ket{\text{TC}} = \prod_v e^{\beta_1P^{(1)}_v + \beta_2P^{(2)}_v}\ket{\text{TC}},
\end{equation}
where the projectors at each vertex $v$ are defined as
\begin{align}
    P^{(1)}_v &= \frac{1}{8}(1+Z_{v(A)}Z_{v(B)})(1+Z_{v(C)}Z_{v(D)})(1+Z_{v(A)}Z_{v(D)}),\\
    P^{(2)}_v &= \frac{1}{8}(1+Z_{v(A)}Z_{v(B)})(1+Z_{v(C)}Z_{v(D)})(1-Z_{v(A)}Z_{v(D)}).
\end{align}
Note that here we use the labeling convention
\begin{equation}
   \includegraphics[scale = 1]{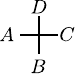}.
\end{equation}
The parameters are related to $g$ via
\begin{align}\label{sm:eq:beta}
    \beta_1 &= \frac{1}{2}\left(\log 2-\log(1+|g|)\right),\\
    \beta_2 &= \frac{1}{2}\left(\log 2-\log(1+|g|)\right) + \frac{1}{2}\log g.
\end{align}
To derive a local parent Hamiltonian, we employ the method in Ref.~\cite{Haller:2023}. Since $(b_p\mathcal{T}b_p\mathcal{T}^{-1}-b_p)\ket{\Psi(g)} = 0$, where $b_p = \prod_{i\in p}X_i$. If follows that
\begin{equation}
    \left(e^{\Lambda_{a}+\Lambda_c +O_b+O_d}-b_p\right)\ket{\Psi(g)} = 0,
\end{equation}
where we use the following labeling convention for the vertices on each plaquette
\begin{equation}
    \includegraphics[scale = 1]{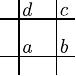}.
\end{equation}
The operators at each vertex are defined as
\begin{align}
    \Lambda_v &= -\frac{1}{4}(\beta_1-\beta_2)\left(1+Z_{v(A)}Z_{v(B)}\right)\left(1+Z_{v(C)}Z_{v(D)}\right)Z_{v(A)}Z_{v(D)},
    \nonumber\\
    O_v &= -\frac{1}{4}\left(Z_{v(A)}Z_{v(B)}+Z_{v(C)}Z_{v(D)}\right)\left[(\beta_1+\beta_2)+(\beta_1-\beta_2)Z_{v(A)}Z_{v(D)}\right].
\end{align}
We can use this observation to obtain a suitable local term in the parent Hamiltonian. It is possible to choose a path slightly deviated from, but continuously connected to Eq.~\eqref{sm:eq:beta} such that the resulting wavefunction Eq.~\eqref{sm:eq:imag} is analytic in $g$ for $g\in (-1,1)$. The definition of the parameters $\beta_1,\beta_2$ can then be extended to $g<0$ using an argument in Ref.~\cite{Haller:2023} based on analytic continuation. The path Eq.~\eqref{sm:eq:beta} is therefore also valid for $g\in [-1,1]$, where we define $\log g = \log |g| +i\pi$ for $g<0$. As a sanity check, it can be verified that the imaginary time evolution at $g = -1$ is analytically continued to a finite-depth local unitary symmetric under the global spin flip, and the wavefunction Eq.~\eqref{sm:eq:imag} at $g = -1$ is the same as the SET isoTNS wavefunction discussed in the main text (up to an overall phase factor). Therefore, all the analysis for $g>0$ can be straightforwardly extended to the case $g<0$.

\begin{figure}[t!]
    \centering
\includegraphics{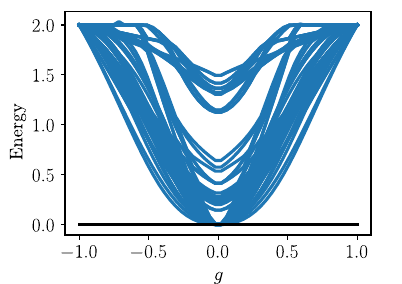}
    \caption{Energy spectrum from the exact diagonalization of a system with $4\times 2$ plaquettes (16 qubits) and periodic boundary condition. The plot shows the low-lying eigenvalues of the Hamiltonian. At $g = \pm 1$, the models are fixed points of the topological phases with an energy gap $\Delta = 2$. }
    \label{sm:fig:energy}
\end{figure}
To proceed, note that
\begin{align}
    &\left(e^{\Lambda_{a}+\Lambda_c +O_b+O_d}-b_p\right)^2 = 
    \nonumber\\
    &\quad 2\cosh(\Lambda_{a}+\Lambda_c +O_b+O_d)\left(e^{\Lambda_{a}+\Lambda_c +O_b+O_d}-b_p\right).
\end{align}
We can therefore define a new plaquette projector as
\begin{align}
    &\mathcal{B}_p(g)P_{p(v)} =\nonumber\\ &\quad \frac{\sech(\Lambda_{a}+\Lambda_c +O_b+O_d)}{2}\left(e^{\Lambda_{a}+\Lambda_c +O_b+O_d}-b_p\right)P_{p(v)}.
\end{align}
where $P_{p(v)}$ is the projector onto the closed loop configuration around each plaquette $p$, i.e. $P_{p(v)} = \prod_{v\in p}(1-\mathcal{A}_v)$. It is included to ensure that $\mathcal{B}_p(g)P_{p(v)}$ remains Hermitian and therefore a projector for $g\in [-1,1]$. The operator hyperbolic secant function is defined as $\sech(O) \equiv 1/\cosh(O)$ within the subspace $P_{p(v)} = 1$, and $\sech(O) \equiv 0$ in the subspace $P_{p(v)} = 0$. A local frustration-free parent Hamiltonian is thus given by
\begin{equation}
      H(g) = \sum_v \mathcal{A}_V + \sum_p  \mathcal{B}_p(g)P_{p(v)},
\end{equation}
with a ground-state energy of zero. The support of each term in $H(g)$ is depicted in Fig.~\ref{sm:fig:parenth}. When the system is defined on a torus (periodic boundary condition), $H(g)$ is gapped and the ground states are exactly four-fold degenerate for $g\neq 0$. At $g = 0$, the system is gapless. In addition,  the $\mathbb{Z}_2^T$ symmetry is satisfied, i.e. $[H(g),\left(\prod_i X_i\right)K] = 0$ for $g\in [-1, 1]$. The low-lying spectrum of the Hamiltonian on a square lattice with $4\times 2$ plaquettes (16 qubits) is shown in Fig.~\ref{sm:fig:energy}.
Note that the parent Hamiltonian for the ground states is not unique, the one derived here is one of the possible parent Hamiltonians.

\section{Finding isoTNS paths between the toric code and the double semion model}\label{sec:sm_tcds}
\begin{figure}[t!]
    \centering
    \includegraphics[width=1\linewidth]{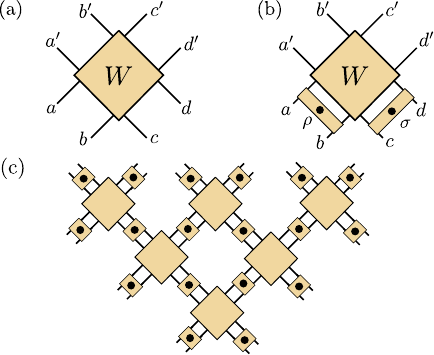}
    \caption{The plumbing construction for the double-line tensor network. (a) We start with a double-line $W$-matrix. (b) The $W$-matrix is plumbed such that the domain walls between the virtual legs are connected to the physical degrees of freedom (black dots) on the links of a square lattice. (c) The full wavefunction is contracted from the plumbed $W$-matrix.}
    \label{fig:sm_tcds}
\end{figure}
The double semion model has a distinct intrinsic topological order from the toric code and supports anyonic excitations with semion statistics~\cite{Freedman:2004, levin:2005}. In this section, we show how to apply the proposed plumbing method to construct an isoTNS path, and therefore an efficient quantum protocol, to realize a topological quantum phase transition between the ground states of the toric code and the double semion model.

We begin by first writing out the isoTNS representation for the fixed-point ground state of the toric code and the double semion model on a square lattice with spin-1/2 degrees of freedom on the links. To achieve this, it is convenient to use the double-line tensor-network representation~\cite{Gu:2008,xu:2018}. Consider a double-line $W$-matrix (see Fig.~\ref{fig:sm_tcds}a). The $W$-matrix has virtual legs of bond dimension 2 and has matrix elements
\begin{equation}
    (W)_{abcd,a'b'c'd'} = A_{acd'b'}\delta_{aa'}\delta_{bc}\delta_{dd'}\delta_{b'c'},
\end{equation}
where $\delta$ denotes the Kronecker delta. Here the tensor $A$ compactly encodes the matrix elements of $W$ when all the Kronecker deltas are satisfied.
As depicted in Fig.~\ref{fig:sm_tcds}b, a tensor-network representation for the double semion ground state is obtained by connecting the domain walls of the $W$-matrix to the physical degrees of freedom via a plumbing delta 
\begin{equation}
    T^{\rho\sigma}_{abcd, a'b'c'd'} = \delta_{ak, bq}^{\rho}\delta_{ch, bl}^{\sigma} W_{kqhl,a'b'c'd'},
\end{equation}
where $\delta_{ak, bq}^{\rho}$ is the plumbing delta for the domain wall such that $\delta_{ak, bq}^{\rho} = 1$ if $a = k, b = q$ and $\rho = a + b\mod 2$. Otherwise the plumbing delta takes a value of zero. The full wavefunction is then obtained from the contraction of the local tensor $T$ (see Fig.~\ref{fig:sm_tcds}c). The resulting TNS has a virtual bond dimension $D = 4$.

Compared to the isometry condition on the single-line TNS as in Eq.~\ref{eq:isometry}, the isometry condition on the double-line TNS involves an additional constraint that the virtual legs $a$ and $c$ in Fig.~\ref{fig:sm_tcds}b have to agree. The additional constraint does not affect the sequential quantum circuit correspondence and the holographic quantum simulation of the isoTNS discussed in the main text---it can be verified that this constraint is automatically satisfied as the tensor network is sequentially contracted from the boundary.
We can derive the condition on the $W$-matrix from the isometry condition as 
\begin{equation}
    \sum_{b',c'}\delta_{b'c'}|W_{abcd,a'b'c'd'}|^2 = \sum_{b'} |A_{acd'b'}|^2 = 1.
\end{equation}
As noted in Ref.~\cite{xu:2018}, the tensor $A$ representing the ground state of the toric code has the elements
\begin{equation}
   A_{acd'b'} = \frac{1}{\sqrt{2}}\text{ for all $a,c,d',b'$.}
\end{equation}
The tensor $A$ representing the ground state of the double semion model has the elements
\begin{equation}
   A_{0011} = A_{0110} = -\frac{1}{\sqrt{2}}\text{ and }A_{acd'b'} = \frac{1}{\sqrt{2}}\text{ otherwise.}
\end{equation}
In both cases, the resulting TNS satisfy the double-line isometry condition and are therefore double-line isoTNS. A simple $D=4$ continuous isoTNS path between the two ground states are given by
\begin{align}
    A(g)_{0010} &= A(g)_{0111}= \frac{1}{\sqrt{1+|g|}},
    \nonumber\\
    A(g)_{0011} &= A(g)_{0110}= \text{sign}(g)\sqrt{\frac{|g|}{1+|g|}},
    \nonumber\\
     A(g)_{1101} &= A(g)_{1000}= \frac{1}{\sqrt{1+|g|}},
    \nonumber\\
    A(g)_{1100} &= A(g)_{1001}= \sqrt{\frac{|g|}{1+|g|}},
\end{align}
and the rest of the elements of the tensor $A(g)$ are 1.
As $g$ is tuned continuously from -1 to +1, the state changes from the ground state of the double semion model to that of the toric code with a distinct intrinsic topological order. This path is similar to the non-isometric TNS path studied in Ref.~\cite{xu:2018}, we therefore expect the similarly unique phase transition point at $g = 0$. At $g = 0$, the wavefunction only contains six allowed closed-loop configurations at each vertex and can be mapped to the classical partition function of the six-vertex model with the same criticality as the example discussed in the main text.

\end{document}